\documentclass[reprint]{revtex4-1}
\usepackage{amsmath,amssymb,graphicx}
\usepackage{graphicx}
\usepackage{dcolumn}
\usepackage{bm}
\usepackage{booktabs}
\usepackage{float}

\begin{document}

\preprint{APS/123-QED}

\title{Improvement in medium-long term frequency stability of integrating sphere cold atom clock}
\author{Peng Liu, Yanling Meng, Jinyin Wan, Xiumei Wang, Yaning Wang,
Ling Xiao}
\author{Huadong Cheng}
\email{corresponding author: chenghd@siom.ac.cn}
\author{Liang Liu}
\email{corresponding author: liang.liu@siom.ac.cn}
 \affiliation{Key Laboratory of Quantum Optics and Center of Cold Atom Physics, Shanghai Institute of Optics and Fine Mechanics, Chinese Academy of Sciences, Shanghai 201800, China.}
\date{\today}

\begin{abstract}
The medium-long term frequency stability of the integrating sphere cold atom clock was improved. During the clock operation, Rb atoms were cooled and manipulated using cooling light diffusely reflected by the inner surface of a microwave cavity in the clock. This light heated the cavity and caused a frequency drift from the resonant frequency of the cavity. Power fluctuations of the cooling light led to atomic density variations in the cavity's central area, which increased the clock frequency instability through a cavity pulling effect. We overcame these limitations with appropriate solutions. A frequency stability of $3.5\times10^{-15}$  was achieved when the integrating time $\tau$  increased to $2\times10^{4}$ s.

\end{abstract}

\maketitle


\section{INTRODUCTION}

The atomic frequency standards have been rapidly developed in the past three decades for applications in space navigation, telecommunication, and timekeeping. In most cases, a portable and stable atom clock is required. Researchers have proposed several schemes for compact atom clocks and attempted to ensure the stability of the atomic clocks for a long operation time. A frequency instability of $3.2\times10^{-15}$   was determined using HORACE when the integration time was  $2\times10^{4}$ s, which shows a good control of system noises  ~\cite{Esnault11}. The pulsed optically pumped (POP) clock developed at INRIM has an instability of $5\times10^{-15}$  at a 4000 s integration time ~\cite{Micalizio15}. The clock based on the intracavity sample of cold Cs atoms proposed by M¨¹ller et al. achieved a frequency stability of $5.0\times10^{-13}\tau^{-1/2}$   and shows a good performance after $10^{4}$ s ~\cite{Muller11}.

The integrating sphere cold atom clock (ISCAC) ~\cite{Liu15} is a kind of compact atomic clock that uses diffuse laser cooling ~\cite{Guillot01,Cheng09,Zhang09} and the POP ~\cite{Godone06pra} operating scheme to achieve a good frequency stability with a small volume and weight. Compared with a vapor cell POP atomic clock, the ISCAC has less atomic collision frequency shifts and less sensitivity to the temperature fluctuations  making it easier to achieve a better frequency stability for a long operation time. Unlike other compact cold atom clocks, ISCAC uses a microwave cavity to diffusely reflect cooling light and cools atoms without a quartz chamber in it ~\cite{Liu15}. This  scheme can eliminate the influence of the quartz chamber on the microwave resonant frequency in the microwave cavity. In addition, the all-metal structure of the clock is more robust and convenient for space applications. However, the cooling-light heating effect on the microwave cavity is considerable and the resonant frequency shift of the microwave cavity should be compensated. The distribution of the cold atoms in the microwave cavity is determined by the injection scheme of the cooling lights and the  corresponding intensities ~\cite{Meng13}. Therefore, in our system, stable cooling lights are necessary for decreasing the cavity pulling effect ~\cite{Bize01}, which is the main limiting factor of the long-term frequency stability.

In this paper, we present the medium-long term frequency stability of ISCAC achieved recently and analyze factors that have a large impact on the frequency stability. The  heating effect on the microwave cavity owing to the cooling lights and the necessity of stabilizing the power of the cooling lights are studied. Section II  briefly introduces the experimental setup and the operating time sequence of the ISCAC. Section III presents the factors limiting of the clock frequency stability and the  solutions  to overcome these limitations. The results of the study of the clock frequency stabilities are presented in this section as well. The final section presents the conclusion and scope of improving the long-term frequency stability of the clock further on.

\section{Experimental setup and time sequence}\label{2}

\begin{figure}[htbp]
\centerline{\includegraphics[width=0.45\textwidth]{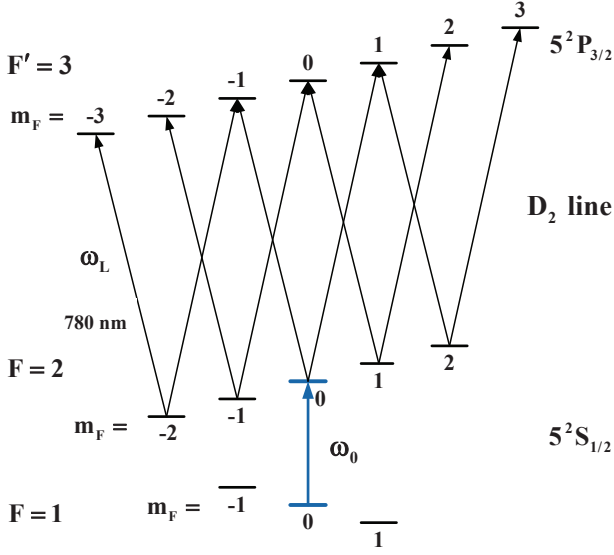}}
\caption{Atomic levels of $^{87}$Rb $D_{2}$ line under biased magnetic field and  transitions during the clock cycles. $\omega_{0}$ and$\omega_{L}$ are the angular microwave and laser frequencies, respectively (colour online).}\label{Fig1}
\end{figure}
The ISCAC is operated using the POP configuration, which involves a diffuse laser cooling process before the optical pumping ~\cite{Liu15}. In the cooling stage, $^{87}$Rb  atoms were cooled to below 100~$\mu$K ~\cite{Wang12} in isotropic cooling light, which is diffusely reflected by the inner surface of the microwave cavity. Then, the atoms were pumped into the $5^{2}S_{1/2}, |F=1\rangle$  state of the  $^{87}$Rb $D_{2}$ line with a short pumping light pulse in preparation for the Ramsey microwave interrogation process. Fig. 1 shows the atomic levels of the $^{87}$Rb $D_{2}$  line and the optical and microwave fields used during the operation. Finally, the atoms that  experienced a clock transition $5^{2}S_{1/2}|F=2,m_{F}=0\rangle\rightarrow 5^{2}S_{1/2}|F=2,m_{F}=0\rangle$ were detected using a standing wave probe light after the second Ramsey pulse. Acousto-optic modulators (AOM) and mechanical shutters were both used to cut off the light to avoid optical frequency shifts of the clock transition. Fig. 2 shows the time sequence of the clock. We discarded the second pumping pulse, which was used for normalizing the  probe light power fluctuation during the detection period ~\cite{Liu15}, because the power lock of the probe light was used instead. In addition, the cycle time without the second pumping pulse was shortened to 88 ms, which reduces the effect of the microwave noise (Dick effect) and simplifies the time sequence. Fig. 3 shows the Ramsey fringes obtained with this time sequence.

\begin{figure}[htbp]
\centerline{\includegraphics[width=0.45\textwidth]{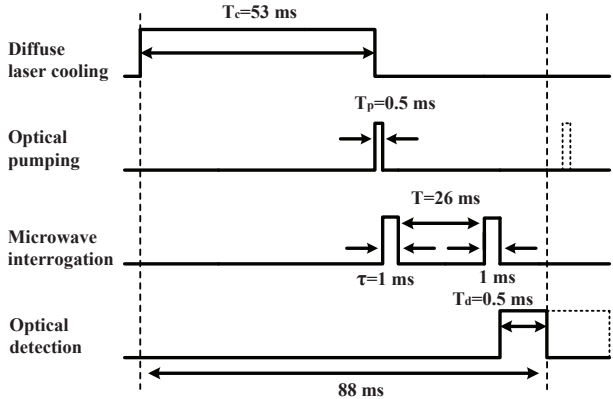}}
\caption{Time sequence of the clock. $T_{c}$ , $T_{p}$,T,$\tau$ and $T_{d}$ are the duration of the cooling, pumping, free evolution, microwave and optical detection pulses, respectively. The dashed lines in the optical pumping and detection stages indicate the time sequence scheme where the second pumping pulse was applied during the detection period. The total cycle time is 88 ms.}\label{Fig2}
\end{figure}
\begin{figure}[htbp]
\centerline{\includegraphics[width=0.48\textwidth]{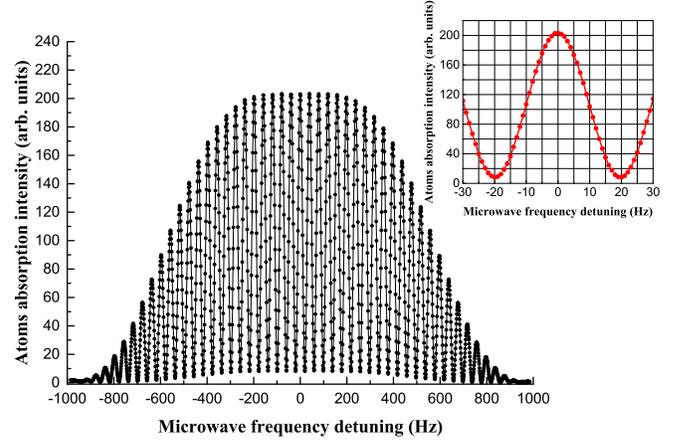}}
\caption{Ramsey fringes obtained with the time sequence shown in Fig.2 and inset is the expanded central fringe with a line-width of 20 Hz (colour online).}\label{Fig3}
\end{figure}
In previous experiments, the error signals for correcting the frequency of the local oscillator (LO) were feed by electrical tuning. A frequency auxiliary output generator, which is phase locked to the LO, was used to decrease the feedback noise in the LO, and its output frequency can be controlled by directly offsetting the phase accumulator in its phase lock loop chain.  The 5 MHz output after correction serves as the reference for the microwave synthesizer, which produces a 6.834 GHz signal. A schematic of the clock and the lock loop is shown in Fig. 4.

\begin{figure}[htbp]
\centerline{\includegraphics[width=0.48\textwidth]{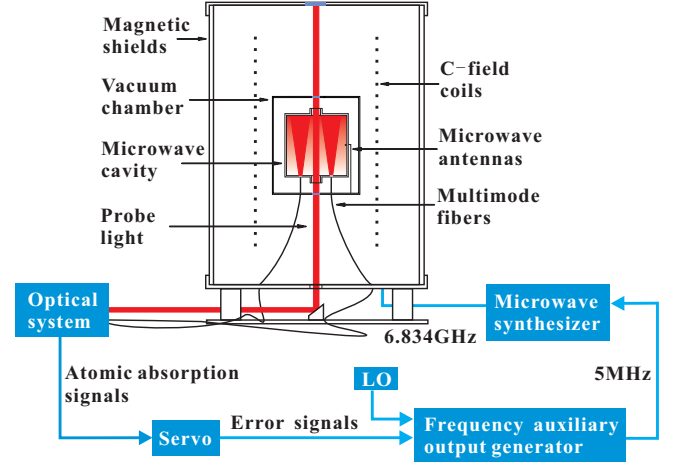}}
\caption{Schematic of clock system (colour online). LO: local crystal oscillator.}\label{Fig4}
\end{figure}

\section{Discussions and results}\label{4}
There are many factors that affect the clock frequency stability. Typically, the shot noise contribution to the frequency stability can be evaluated theoretically using the signal to noise ratio (R$_{SN}$) and quality factor (Q$_{a}$) of the atomic resonance line ~\cite{Vanier89}. The Dick effect contribution to frequency stability can be evaluated with the frequency noise spectrum of the LO and the duty cycle of the microwave interrogation process ~\cite{Santarelli98}. For the ISCAC, the Dick effect contribution decreased from $3.3\times10^{-13}\tau^{-1/2}$ to $2.4\times10^{-13}\tau^{-1/2}$ after adjusting the time sequence. However, there are other factors, such as temperature fluctuations of the microwave cavity and variation in the biased magnetic field, that limit on the medium-long term frequency stability. Therefore, to achieve better clock frequency stability, several physical and technical problems have to be solved.
\begin{figure}[htbp]
\centerline{\includegraphics[width=0.45\textwidth]{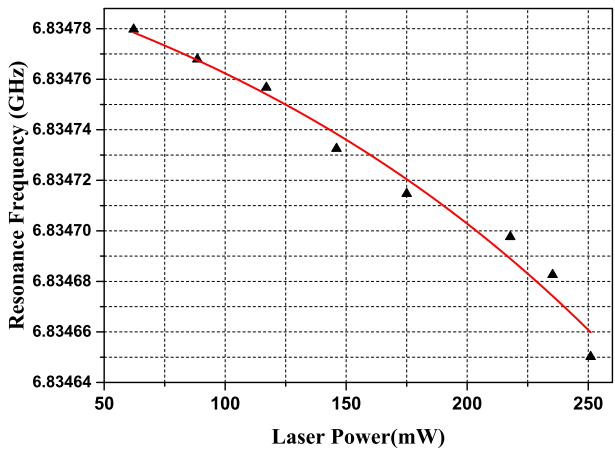}}
\caption{Relationship between cooling laser power and the microwave cavity resonance frequency. The black triangles are the experimental data while the red curve is the fitting result with an exponential function (colour online).}\label{Fig5}
\end{figure}

Temperature stability is a very important factor that affects the stability of an atomic clock because many physical characteristics of the clock devices are temperature dependent.  For the ISCAC, because the resonant frequency of the cylindrical microwave cavity is susceptible to temperature variations, we implemented a two stage temperature control loop to stabilize the microwave cavity temperature. However, this measure is not enough to ensure temperature stabilization, especially after a long period of operation . Owing to the special cooling scheme of the ISCAC, the cooling light, which has a total power about 200 mW was directly reflected by the microwave cavity with the inner diameter and height of 57 mm in each cycle. The laser heating effect on the microwave cavity is not negligible. Fig. 5 shows the relationship between the injected laser power and the microwave cavity resonant frequency. When we adjust the resonant frequency, the detuning caused by the heating effect must be considered and compensated by the temperature control loop. In addition, the cold atom distribution in the microwave cavity is largely determined by the cooling light ~\cite{Meng13}. Power fluctuations in the cooling lights will transfer to the instabilities of the cold atoms in the microwave cavity. The fluctuations in the number of cold atoms  will further degrade the clock frequency stability  according to the model for estimating the cavity frequency pulling effect of an atomic fountain ~\cite{Bize01}:
\begin{equation}
\Delta\nu_{cp}\propto\frac{\tau}{T}\frac{\mu_{0}\mu_{B}^{2} N_{i} Q}{2\pi^{2}\hbar V_{mode}}[\frac{2(\omega_{a}^{2}/Q)(\omega_{a}^{2}-\omega_{c}^{2})}{(\omega_{a}^{2}-\omega_{c}^{2})^{2}+(\omega_{a}\omega_{c}/Q)^{2}}],
\label{Delta}
\end{equation}
where Q and $V_{mode}$ are the TE011 mode quality factor and mode volume,respectively. $\omega_{c}/2\pi$   is the cavity eigenmode resonance frequency and $\hbar\omega_{a}$ is the clock transition energy.  $N_{i}$ is the number of cold atoms that have experienced the microwave interrogation process.In our case, cavity pulling effect is a dominant factor to affect the medium-long term stability because the whole stage of Ramsey interrogation is carried on in the microwave cavity~\cite{Rossetto11}. Therefore, the intensity stability of cooling light has a larger impact on the clock frequency stability than other schemes of cold atom clocks. An active power stabilization loop was required to decrease the cooling light fluctuations. After this servo loop was implemented, the temperature stability of the microwave cavity was found to not be a restrictive factor affecting the clock frequency stability. The temperature stability of the microwave cavity during the experiment is shown in Fig. 6. The temperature sensitivity of the microwave cavity resonant frequency was 113.5 KHz/$^{\circ}$C , which was calibrated in a vacuum environment before the installation ~\cite{Meng14}. Then the relative resonant frequency variation $\delta\nu_{c}/\nu_{c}$  was less than $2.0\times10^{-8}$ and the contribution to the clock frequency instability of the microwave cavity temperature fluctuations is at the level of $2.6\times10^{-16}$.
\begin{figure}[htbp]
\centerline{\includegraphics[width=0.45\textwidth]{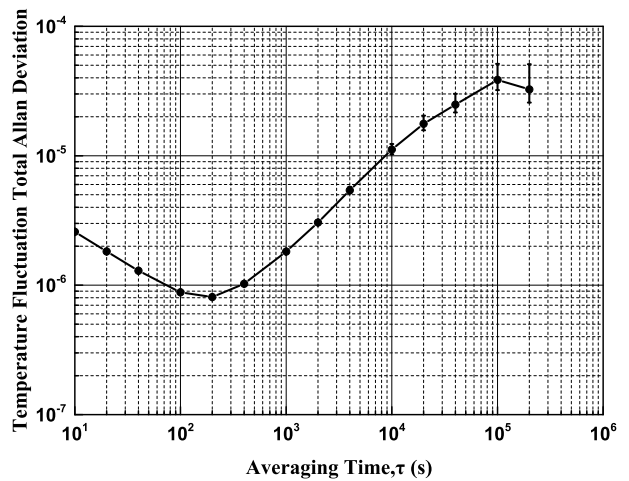}}
\caption{The temperature fluctuation of the microwave cavity.}\label{Fig6}
\end{figure}
The microwave cavity is covered with five layers of magnetic shields which provide a well-shielded environment. The quantization magnetic field (C-field) is driven by two low noise current sources. Fig. 7 shows the measured stability of the current source. the second-order Zeeman shift of the clock frequency can be estimated using the equation shown below ~\cite{Vanier89}:

\begin{equation}
\Delta\nu_{z}=K_{0}|\mathbf{B}\cdot\hat{z}|^{2},
\label{Delta1}
\end{equation}
where $K_{0}=575.14\times10^{8}Hz/T^{2}$,$\mathbf{B}$ is the biased magnetic field intensity and $\hat{z}$ is the unit factor of the quantization axis. To reduce the effect of the second-order Zeeman shift on the clock frequency stability, we reduced the  the C-field intensity to 3.1 mG. In this situation, the second-order Zeeman shift was ~ $5.5\times10^{-6}$Hz  and the clock instability degraded to $3.2\times10^{-17}$ when the average time increased to $2\times10^{4}$ s, which can be ignored for the system.
\begin{figure}[htbp]
\centerline{\includegraphics[width=0.45\textwidth]{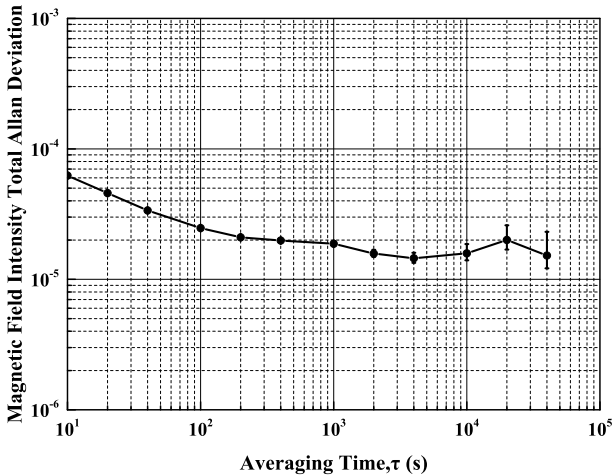}}
\caption{Stability of biased magnetic field current source.}\label{Fig7}
\end{figure}

Typically, the frequency stability of the ISCAC is measured by comparing it with an H-maser whose stability is $2.0\times10^{-13}$  at 1 s and $2.0\times10^{-15}$ after $10^{4}$ s. Fig. 8 shows two sets of frequency stability results of the clock with different experimental conditions. The Allan deviation (black triangles) result is measured without locking the cooling light power and the cycle time is 133 ms. The red dots curve is the clock frequency stability obtained after applying cooling light power stabilization and the clock cycle time is shortened to 88 ms. The white frequency region is extended  to $2\times10^{4}$ s and a frequency stability of $3.5\times10^{-15}$  was achieved at $2\times10^{4}$ s after the improving the experimental conditions.
\bigskip
\begin{figure}[H]
\centerline{\includegraphics[width=0.45\textwidth]{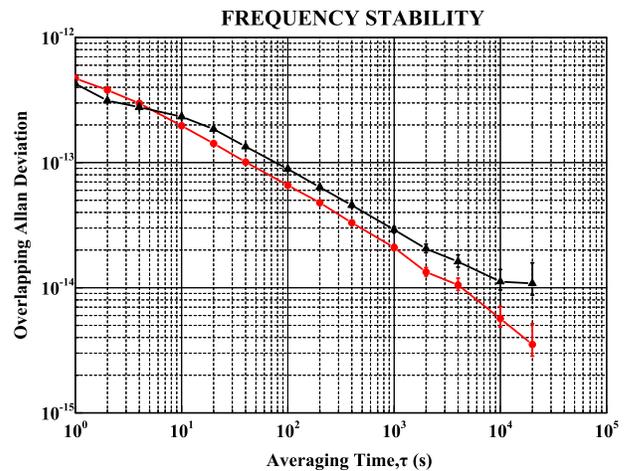}}
\caption{Frequency stabilities of ISCAC under medium-term measurements (colour online).}\label{Fig8}
\end{figure}

\section{CONCLUSIONS}\label{5}
We reported the improvement on the medium-long term frequency stability of the ISCAC . The systematic factors that affect the clock medium-long term frequency stability are discussed and their limitations on the clock frequency stability were reduced by improving the experimental conditions. In addition, the laser heating effect on the microwave cavity and the effect of the cooling laser power fluctuations were studied and restrained by employing appropriate solutions. The clock frequency stability reached less than $10^{-15}$  when  $\tau>$4000 s and $3.5\times10^{-15}$ at $2\times10^{4}$s. A new microwave synthesizer of low phase noise and small volume is in the development to decrease the noise and instability of the microwave interrogation pulses.  Meanwhile, further investigation on improving the clock long-term frequency stability performance will be conducted to determine the requirements for practical applications.

\end{document}